\documentclass[article]{revtex4}
\usepackage{graphicx}
\usepackage{amssymb}
\usepackage{amsmath}
\usepackage{bm}
\usepackage{hyperref}
\usepackage{epstopdf}

\begin{document}

\title{Generalized Noether Theorem for Gauss-Bonnet Cosmology}
\author{Han Dong $^{1}$}\email{donghan@mail.nankai.edu.cn}

\affiliation{$^{1}$ Fuzhou Institute of Technology, Fuzhou 350506, China}

\date{\today}

\begin{abstract}
Generalized Noether's theory is a useful method for researching the modified gravity theories about the conserved quantities and symmetries. A generally Gauss-Bonnet gravity $f(R,\mathcal{G})$ theory was proposed as an alternative gravity model. Through the generalized Noether symmetry, polynomial and product forms of the $f(R,\mathcal{G})$ theory with corresponding conserved quantities and symmetries are researched. Then suitable general forms of the polynomial form $f(R,\mathcal{G}) \!=\! k_1 R^n + (6)^{\frac{n}{2}}(-1)^{n+1} k_1 \mathcal{G}^{\frac{n}{2}}$ and the product form $f(R,\mathcal{G}) \!=\! k ( R / \sqrt{\mathcal{G}} )^n  \mathcal{G}$ are found out, to contain the solution of accelerated expansion cosmology. Both forms of $f(R,\mathcal{G})$ concerned in this paper only possess time translational symmetry. And energy condition of these solutions are also checked. To some extent, the consistency of conservation of symmetry and energy condition is demonstrated. For the specific form of different $n$, it needs further detailed study. Noting that, the corresponding conserved quantities are both zero, and the only conservation relation is conservation of energy.
\end{abstract}


\maketitle

\section{Introduction}

Accelerated expanding of our universe has been widely known, which is confirmed by various sets of astrophysical observations including WAMP~\cite{WAMP}, LSS~\cite{LSS}, SN Ia~\cite{SNe Ia} and SDSS~\cite{SDSS}. These observations also imply that our universe consists of about 68\% dark energy (DE), 32\% dust matter, and negligible relativistic constituent at present, and is spatially flat on large scales. For explaining our universe, the standard cosmological models have been proposed, however some problems describing the whole universe still exist, such as initial singularity, flatness problem, and so on. Moreover, the general relativity can not be used as a fundamental theory to describe the quantum properties of space-time. Because of the above shortcomings, people try to introduce gravitational quantization into the modification of general relativity. For this reason, the $f(R,{\cal G})$ theory is proposed to obtain an effective reconfigurable quantum gravity theory at Planck scale. The Ricci scalar $R$ and the Gauss-Bonnet topological invariant ${\cal G}$ are added to the general relativity~\cite{effective action2,OdintsovPR1,OdintsovPR2,F(GR)-gravity,antonio}. These theories are stable and able to describe the present accelerated expanding of universe. Through manual selected concrete form of the $f(R,{\cal G})$ theory, people can rebuild models to mimic the $\Lambda$CDM model\cite{diego,myrz,diego3,Felice_malo}.

Since there are no fundamental constraints, most of the models generated by the theory of gravitational expansion in the past researches are proposed from phenomenology, which seems to be somewhat unconvincing. The explicit expressions of Extended Theories of Gravity will determine the exact forms of solutions of the field equation. It is important to analyze physical properties of the theories. In this paper, our concern is whether there are some basic principles that can provide reasonable and acceptable modifications.

Fortunately, the famous Noether's theorem is suitable for the purposes mentioned above, and it has been extensively studied to some models, ie., in scalar field cosmology~\cite{scalar_Noether}, non-minimally coupled cosmology~\cite{non-minimally_coupled_Noether}, $f(R)$ theory~\cite{fR_Noether}, $f(T)$ theory~\cite{fT_Noether,fT_birkhoff}, $f(R,\mathcal{G})$ theory~\cite{fGBnoether,fRGBnoether}, $f(T,\mathcal{G})$ theory~\cite{fTGBnoether}, scalar-tensor theory~\cite{scalar-tensor_Noether} and quantum cosmology~\cite{quantum_Noether}. Compared with Noether's theorem, it has been extended to a more general generalized Noether's theorem for the complete solutions~\cite{Noether_symmetry,General_Noether_symmetry1,General_Noether_symmetry2}. Using the generalized Noether symmetry, the research has also been extensively done, ie., in $f(R)$ theory~\cite{fRGN}, and $f(T)$ theory~\cite{fTGN,dis Noether}. In our previous paper~\cite{dis Noether}, it has been proved that the approach of generalized Noether symmetry is more physical content, such as the energy condition. Moreover, this approach is a more effective choice method, a more complete solutions, and more conformed to cosmological observations. By this generalized Noether symmetry approach, we can find the physically interesting forms of modified gravity theories, and the existence of symmetries allowed to select constants of motion that reduce dynamics. Physical contents and the relations to cosmological or astrophysical observations of the conserved quantities are very interesting.

Most of the previous papers, the researches were limited to Noether symmetries, whose the coefficients of vector fields were set to depend only on coordinates and time in configuration space. The generalization of Noether's theorem was realized by adding the first and higher order time derivatives of coordinates into the coefficient functions of vector fields. Therefore researchers can get the complete set of solutions, like the generalized Noether symmetries and the corresponding generalized vector fields \cite{General_Noether_symmetry1}. To gain insight of this theorem, we should consider the motion of a particle under the influence of a central force (Kepler situation), characterized by the Lagrangian
\begin{equation}
L={1\over 2}\left[\left(\dot{r}^2+r^2\dot{\theta^2}\right)+{k\over
r}\right].
\end{equation}
In this situation, the generalized vector fields are \cite{General_Noether_symmetry2}
\begin{eqnarray}
X_1&=&r^2\cos\theta\dot{\theta}{\partial\over \partial
r}+\left(\cos\theta\dot{r}-2r\sin\theta\dot{\theta}\right){\partial\over
\partial \theta}, \nonumber \\
X_2&=&r^2\sin\theta\dot{\theta}{\partial\over \partial
r}+\left(\sin\theta\dot{r}+2r\cos\theta\dot{\theta}\right){\partial\over
\partial \theta}.
\end{eqnarray}
They are just variational symmetries of lagrangian related to Runge-Lenz vector \cite{General_Noether_symmetry1} of the perihelia precession. This solution is not available by the Noether symmetry. Obviously, the complete solutions can only be obtained via the generalized Noether symmetry, rather than Noether symmetry.

In this paper, we study in detail some manifestations of the $f(R,{\cal G})$ gravity through the generalized Noether theorem.
We try to constrain the functional form of $f(R,{\cal G})$, and get the exact late-time accelerated expanding solutions as well as the corresponding
conserved quantities. This paper is arranged as follows:
In Sec. \ref{tre}, the Gauss-Bonnet gravity will be briefly introduced. In Sec. \ref{noether}, we thoroughly research the generalized Noether symmetries of $f(R,{\cal G})$ manifestations, ie.,$f(R,\mathcal{G}) \!=\! k_1 R^n + k_2 \mathcal{G}^m$ and $f(R,\mathcal{G}) \!=\! k R^n \mathcal{G}^m$.
In Sec. \ref{conclusion}, we will sum up some conclusions through discussion.

\section{Brief Introductory to Gauss-Bonnet Cosmology}
\label{tre}

In 4-dimensions spacetime, the most general action of extended Gauss-Bonnet gravity is shown as
\begin{equation}
{\cal S}=\frac{1}{2\kappa}\int d^4x \sqrt{-g} \,f(R,{\cal G})\,,
   \label{action1}
\end{equation}
where we define $\kappa^2\!=\!8\pi G_N$, $G_N$ as the Newton constant, $g$ as the determinant of the metric as well as the physical units $c \!=\! k_B \!=\! \hbar \!=\!1$.
This Lagrangian is constructed only by the metric tensor, without any extra vector or spin degree of freedom.
The Gauss-Bonnet invariant is defined as
\begin{equation}
 {\cal G}\,\equiv\,R^2-4R_{\mu\nu}R^{\mu\nu}+R_{\mu\nu\lambda\sigma}R^{\mu\nu\lambda\sigma}\,,
\label{GB}
\end{equation}
that is a combination of the Ricci scalar $R\!=\!g^{\mu\nu}R_{\mu\nu}$, the Ricci tensor $R_{\mu\nu}$ and the Riemann tensor $R_{\mu\nu\lambda\sigma}$. Noting that, in 4-dimension, any linear combination of the Gauss-Bonnet invariant does not
contribute to the effective Lagrangian.
The variation of the action (\ref{action1}) with respect to the metric provides the following gravitational field equation \cite{antonio}
\begin{eqnarray}\label{eom}
G_{\mu \nu} &=& \frac{1}{ f_R} \Biggr[\nabla_\mu \nabla_\nu f_{R}-g_{\mu \nu} \Box f_{R}+2R \nabla_\mu \nabla_\nu f_{\cal G} - 4R_\mu^{~\lambda} \nabla_\lambda \nabla_\nu f_{\cal G} - 4R_\nu^{~\lambda} \nabla_\lambda \nabla_\mu f_{\cal G} + 4R_{\mu \nu} \Box f_{\cal G} \nonumber \\
&+& 4 g_{\mu \nu} R^{\alpha \beta} \nabla_\alpha \nabla_\beta f_{\cal G} + 4R_{\mu \alpha \beta \nu} \nabla^\alpha \nabla^\beta f_{\cal G} - 2g_{\mu \nu} R \Box f_{\cal G} - \frac{1}{2}\,g_{\mu \nu}\bigr( R  f_R + {\cal G}  f_{\cal G}- f(R,{\cal G})\bigr) \Biggr] .
\end{eqnarray}
The trace is
\begin{equation} \label{trace}
3\left[\Box f_R + V_R\right]+R \left[\Box f_{\cal G}+ W_{\cal G}\right]=0,
\end{equation}
where $\Box$ is the d'Alembert operator in curved spacetime and
\begin{equation}
  \label{eq:def1}
  f_{R}\equiv\frac{\partial f(R,{\cal G})}{\partial R}\,,\qquad f_{\cal G}\equiv \frac{\partial f(R,{\cal G})}{\partial {\cal G}}\,,
\end{equation}
are the partial derivatives with respect to $R$ and $\cal G$.
Obviously, if we assume $f(R,{\cal G})=R$, General Relativity will be immediately recovered from Eqs.~(\ref{eom})-(\ref{trace}).

\section{Generalized Noether Symmetry for Gauss-Bonnet Cosmology}
\label{noether}

In 1918, Noether proposed an elegant and systematic approach to compute conserved quantities for the partial differential equations (PDEs) by a systematic and mathematical way. The field equations in modified gravity are just the PDEs, and the solutions can be investigated by the Noether's theorem. The Noether's theorem states that each conservation law corresponds to a differentiable symmetry of the action of a physical system.
Then the conservation laws play an important role in the research of physical phenomena. For getting a more complete solution of the system, the configuration space of the generalized Noether symmetries is extended to the first or higher order of time derivatives of coordinates \cite{General_Noether_symmetry1}, and the symmetries are defined as transformations in the configuration space.

We choose the flat FRW space-time, whose signature is $(+,-,-,-)$, and the explicit descriptions of Ricci scalar and Gauss-Bonnet invariant are shown as
\begin{equation}
    R = -6\,(\frac{\ddot{a}}{a}+\frac{\dot{a}^2} {a^2}),\qquad   \mathcal{G} = 24\,(\frac{\ddot{a}\, \dot{a}^2}{a^3}).
\end{equation}
The method of Lagrange multipliers is done by setting $R$ and $\mathcal{G}$ constraints of the dynamics, then we naturally get the point-like FRW Lagrangian ${\cal L}(t, a, {\dot a}, R, {\dot R}, {\cal G}, {\dot{\cal G}})$
\begin{equation}
    {\cal L} = a^{3}\bigg(f - f_{R}\bigg[R+6\,(\frac{\ddot{a}}{a}+\frac{\dot{a}^2} {a^2})\bigg] - f_{\mathcal{G}}\bigg[\mathcal{G}- 24\,(\frac{\ddot{a}\, \dot{a}^2}{a^3})\bigg]\bigg).
\end{equation}
By partial integration, the point-like Lagrangian changes as the following form
\begin{eqnarray}\label{Lagrangian f(R,G)}
{\cal L} &=& a^{3}(f-Rf_{R}-\mathcal{G}f_{\mathcal{G}}) + 6\, a \dot{a}^2 f_{R} +  6\, a^2 \dot{a}(\dot{R}f_{RR}+\dot{\mathcal{G}}f_{R\mathcal{G}})-8\,\dot{a}^3(\dot{\mathcal{G}}f_{\mathcal{GG}}+\dot{R}f_{\mathcal{G}R}).
\end{eqnarray}
${\cal Q}\equiv\{t,a,R,\mathcal{G}\}$ is the configuration space, therefore the generalized Noether symmetric generator is shown as
\begin{eqnarray}\label{symmetry generator f(R,G)}
    X &=& \tau(t,a,R,\mathcal{G})\frac{\partial}{\partial t} + \alpha(t,a,R,\mathcal{G})\frac{\partial}{\partial a} + \beta(t,a,R,\mathcal{G})\frac{\partial}{\partial R} + \gamma(t,a,R,\mathcal{G})\frac{\partial}{\partial \mathcal{G}},
\end{eqnarray}
where $\tau$, $\alpha$, $\beta$ and $\gamma$ are the smooth functions of the independent variables $t$ and the canonical coordinates $a$, $R$ and $\mathcal{G}$. So the first prolongation $X_{[1]}$ can be shown as
\begin{equation}\label{first prolongation}
    X_{[1]} = X + \dot{\alpha}\frac{\partial}{\partial \dot{a}} + \dot{\beta}\frac{\partial}{\partial \dot{R}} + \dot{\gamma}\frac{\partial}{\partial \dot{\mathcal{G}}}.
\end{equation}
In addition, the Lagrangian quantity will satisfy the following equation:
\begin{equation}\label{Noether equation f(R,G)}
    X_{[1]} {\cal L} + (D\;\tau) {\cal L} = D\; B(t,a,R,\mathcal{G}),
\end{equation}
where $B(t,a,R,\mathcal{G})$ indicates the gauge function, and $D$ is the total derivative.
The first integral is defined as the time integral of the conserved quantity corresponding to $X$, as shown below
\begin{equation}\label{conserved quantity f(R,G)}
    I = B - \tau {\cal L} - (\alpha - \tau \dot{a})\frac{\partial {\cal L}}{\partial \dot{a}} - (\beta - \tau \dot{R})\frac{\partial {\cal L}}{\partial \dot{R}} - (\gamma - \tau \dot{\mathcal{G}})\frac{\partial {\cal L}}{\partial \dot{\mathcal{G}}}.
\end{equation}
On the other hand, the energy equation for a point-like canonical Lagrangian is
\begin{equation}\label{energy equation}
  E_{\cal L} = \frac{\partial {\cal L}}{\partial \dot{a}} \dot{a} + \frac{\partial {\cal L}}{\partial \dot{R}} \dot{R} + \frac{\partial {\cal L}}{\partial \dot{\mathcal{G}}} \dot{\mathcal{G}} -{\cal L} = 0.
\end{equation}
If we consider the energy equation into the conserved quantity $I$, the conserved quantity $I$ can be reduced to
\begin{equation}\label{conserved quantity and energy equation}
    I = B - \alpha \frac{\partial {\cal L}}{\partial \dot{a}} - \beta \frac{\partial {\cal L}}{\partial \dot{R}} - \gamma \frac{\partial {\cal L}}{\partial \dot{\mathcal{G}}}.
\end{equation}
Taking the point-like Lagrangian~(\ref{Lagrangian f(R,G)}) of $f(R,\mathcal{G})$ into Eq.~(\ref{Noether equation f(R,G)}), we can can obtain a determined system of linear partial differential equations, according to $\dot{a}, \dot{R}, \dot{\mathcal{G}}, \dot{a}\dot{R}, \dot{a}\dot{\mathcal{G}}, \dot{R}\dot{\mathcal{G}}$ and so on. In this general research, we obtain twenty-seven partial differential equations:
\begin{eqnarray} \label{PDEs1 f(R,G)}
  0 &=& \alpha f_{R} + \beta a f_{RR} + \gamma a f_{R\mathcal{G}} + \tau_{t} a f_{R} + 2 a \alpha_{a} f_{R} + a^2 \beta_{a} f_{RR} + a^2 \gamma_{a} f_{R\mathcal{G}} , \nonumber \\
  0 &=& 2\alpha f_{RR} + \beta a f_{RRR} + \gamma a f_{RR\mathcal{G}} + a \alpha_{a} f_{RR}  + 2 \alpha_{R} f_{R} + a \beta_{R} f_{RR} + a \gamma_{R} f_{R\mathcal{G}} + \tau_{t} a f_{RR}, \nonumber \\
  0 &=& 2 \alpha f_{R\mathcal{G}} + \beta a f_{\mathcal{G}RR} + \gamma a f_{\mathcal{GG}R} + a \alpha_{a} f_{R\mathcal{G}} + 2 \alpha_{\mathcal{G}} f_{R} + a \beta_{\mathcal{G}} f_{RR} + a \gamma_{\mathcal{G}} f_{R\mathcal{G}} + \tau_{t} a f_{R\mathcal{G}}, \nonumber \\
  0 &=& \alpha_{R} f_{R\mathcal{G}} + \alpha_{\mathcal{G}} f_{RR}, \nonumber \\
  0 &=& \alpha_{R} f_{RR}, \nonumber \\
  0 &=& \alpha_{\mathcal{G}} f_{R\mathcal{G}}, \nonumber \\
  0 &=& \tau_{a} 6 a^2 f_{RR} + \tau_{R} 6 a f_{R}, \nonumber \\
  0 &=& \tau_{a} 6 a^2 f_{R\mathcal{G}} + \tau_{\mathcal{G}} 6 a f_{R}, \nonumber \\
  0 &=& \tau_{R} 6 a^2 f_{R\mathcal{G}} + \tau_{\mathcal{G}} 6 a^2 f_{RR}, \nonumber \\
  0 &=& \tau_{a} 6 a f_{R}, \nonumber \\
  0 &=& \tau_{R} 6 a^2 f_{RR}, \nonumber \\
  0 &=& \tau_{\mathcal{G}} 6 a^2 f_{R\mathcal{G}}, \nonumber \\
  0 &=& \beta f_{\mathcal{G}RR} + \gamma f_{\mathcal{GG}R} + \beta_{R} f_{R\mathcal{G}} + \gamma_{R} f_{\mathcal{GG}} + 3\alpha_{a} f_{R\mathcal{G}} + \tau_{t} f_{\mathcal{G}R}, \nonumber \\
  0 &=& \beta f_{\mathcal{GG}R} + \gamma f_{\mathcal{GGG}} + \beta_{\mathcal{G}} f_{R\mathcal{G}} + \gamma_{a} f_{\mathcal{GG}} + 3\alpha_{a} f_{\mathcal{GG}} + \tau_{t} f_{\mathcal{GG}}, \nonumber \\
  0 &=& \alpha_{\mathcal{G}} f_{\mathcal{GG}}, \nonumber \\
  0 &=& \alpha_{R} f_{R\mathcal{G}}, \nonumber \\
  0 &=& \alpha_{R} f_{\mathcal{GG}} + \alpha_{\mathcal{G}} f_{R\mathcal{G}}, \nonumber \\
  0 &=& \beta_{a} f_{R\mathcal{G}} + \gamma_{a} f_{\mathcal{GG}}, \nonumber \\
  0 &=& -8 \tau_{a} f_{\mathcal{GG}}, \nonumber \\
  0 &=& -8 \tau_{a} f_{\mathcal{G}R}, \nonumber \\
  0 &=& -8 \tau_{R} f_{\mathcal{GG}} - 8\tau_{\mathcal{G}} f_{\mathcal{G}R}, \nonumber \\
  0 &=& -8 \tau_{\mathcal{G}} f_{\mathcal{GG}}, \nonumber \\
  0 &=& -8 \tau_{R} f_{\mathcal{G}R},  \nonumber \\
  B_{a} &=& \tau_{a} a^3 (f-Rf_{R}-\mathcal{G}f_{\mathcal{G}}), \nonumber \\
  B_{R} &=& \tau_{R} a^3 (f-Rf_{R}-\mathcal{G}f_{\mathcal{G}}), \nonumber \\
  B_{\mathcal{G}} &=& \tau_{\mathcal{G}} a^3 (f-f_{R}R-f_{\mathcal{G}}\mathcal{G}), \nonumber \\
  B_{t} &=& 3 \alpha a^2 (f-Rf_{R}-\mathcal{G}f_{\mathcal{G}}) - \beta a^3 (Rf_{RR} + \mathcal{G}f_{\mathcal{G}R}) - \gamma a^3 (\mathcal{G}f_{\mathcal{GG}} + Rf_{R\mathcal{G}}) + \tau_{t} a^3 (f-Rf_{R}-\mathcal{G}f_{\mathcal{G}}). \nonumber \\
\end{eqnarray}
We could find that if we assumed $\tau\!=\!0$, the determined system of linear PDEs should degenerate to the case of Noether symmetry~\cite{fGBnoether}, and naturally have the same corresponding Noether vector, symmetric generators and solutions. Moreover,
we also can find a special solution of the $B\!=\!0$ from the last four formulas of the Eqs.~(\ref{PDEs1 f(R,G)}), for a special condition
\begin{equation}\label{special condition}
    \alpha = \beta = \gamma = 0 \quad and \quad \tau = const.
\end{equation}
So the Eq.~(\ref{conserved quantity f(R,G)}) changes as
\begin{equation}
    I = const \cdot \bigg(\frac{\partial {\cal L}}{\partial \dot{a}} \dot{a} + \frac{\partial {\cal L}}{\partial \dot{R}} \dot{R} + \frac{\partial {\cal L}}{\partial \dot{\mathcal{G}}} \dot{\mathcal{G}} -{\cal L} \bigg) = const \cdot E_{\cal L} = const \cdot (p_{i} u^{i} - L) = 0.
\end{equation}
If we can obtain this special condition (\ref{special condition}), then we can get the energy equation (\ref{energy equation}). This means that there is a relation between the conserved quantity $I\!=\!0$ and the energy condition.

Next, we will analyze in detail two general forms (models) of manifestation of $f(R,\mathcal{G})$ theory with corresponding generalized Noether symmetries rather than Noether symmetries, cosmological solutions of each model. Firstly, we don't assume any special condition of the last equation of system~(\ref{PDEs1 f(R,G)}) to get the chosen functional forms~\cite{fGBnoether}. In other word, we consider two general forms for simplicity, expressed as
\begin{eqnarray}
  f(R,\mathcal{G}) &=&  k_1 R^n + k_2 \mathcal{G}^m ,  \nonumber\\
  f(R,\mathcal{G}) &=&  k R^n \mathcal{G}^m.  \nonumber\\
\end{eqnarray}

\subsection{Case \uppercase\expandafter{\romannumeral1}: $f(R,\mathcal{G}) = k_1 R^n + k_2 \mathcal{G}^m$}

For this form $f(R,\mathcal{G}) \!=\! k_1 R^n + k_2 \mathcal{G}^m$, the simplest and most commonly considered expression is adopted. According to the previous introduction, expanding the determined system of linear partial differential equations~(\ref{PDEs1 f(R,G)}) with this form, the solutions of $\tau$, $\alpha$, $\beta$ $\gamma$ and $B$ are as follows:
\begin{eqnarray}
  \tau &=& -3 F_1(t) + C_{1},  \nonumber\\
  \alpha &=& \dot{F_1}(t) a,  \nonumber\\
  \beta &=& 0,  \nonumber\\
  \gamma &=& 0,  \nonumber\\
  B &=& const,
\end{eqnarray}
where the $C_{i}$ are the normalized orthogonal coefficients, and $F_1(t)$ is the function with time only. Assuming one of the $C_{i}$ to $1$ and the others to $0$, the corresponding generalized Noether symmetric generators are obtained as
\begin{eqnarray}
  X_{1} &=& \bigg(-3 F_1(t) + 1 \bigg)\frac{\partial}{\partial t} + \dot{F_1}(t) a\frac{\partial}{\partial a}, \\
  X_{2} &=& -3 F_1(t)\frac{\partial}{\partial t} + \dot{F_1}(t) a\frac{\partial}{\partial a}.
\end{eqnarray}
As we all know, if a system is complete, then its symmetric generators should satisfy the commutation relation.
So we can use the non-vanishing Lie bracket $[X_1, X_2]$ to represent the commutator relations of the set $\{X_1, X_2\}$,
\begin{equation}
    [ X_1 ,  X_2 ]  =  \bigg(-3 \dot{F_1}(t) \bigg)\frac{\partial}{\partial t} + \bigg(\ddot{F_1}(t) a + \dot{F_1}(t)\dot{a} \bigg)\frac{\partial}{\partial a},
\end{equation}
thus the solution must be $X_1$, $X_2$ or $0$ for a complete system.
\begin{eqnarray}
    \bigg(\ddot{F_1}(t) a + \dot{F_1}(t)\dot{a} \bigg)\frac{\partial}{\partial a} &=& \dot{F_1}(t) a\frac{\partial}{\partial a},
\end{eqnarray}
for two terms of $a$ and $\dot{a}$, they must be
\begin{equation}
    \ddot{F_1}(t) = \dot{F_1}(t) = 0,
\end{equation}
so the only remaining generalized noether symmetric generator and the corresponding first integral
\begin{eqnarray}
    X &=& \frac{\partial}{\partial t}, \\
    I &=& 6 a \dot{a}^2 k_{1} R^{n-1} n + a^3 k_{1} R^n (n-1) \bigg( 1 + 6 n \frac{H \dot{R}}{R^2} \bigg) + a^3 k_{2} \mathcal{G}^m (m-1)\bigg( 1 - 24 m \frac{H^3 \dot{\mathcal{G}}}{\mathcal{G}^2} \bigg) \label{R^n+G^m},
\end{eqnarray}
which apparently indicate energy conservation in the universe, but it's difficult to calculate the conserved quantity of arbitrary $n$ and $m$.
In spite of this, we can still analyze the conditions for the de Sitter solution $a(t)\! \propto \! e^{H_{0}t}$. Furthermore, the Ricci scalar $R$ and Gauss-Bonnet invariant $\mathcal{G}$ are transformed into
\begin{equation}
     R = -12 H_{0}^2 \quad and \quad  \mathcal{G} = 24 H_{0}^4,
\end{equation}
which mean that $\dot{R}\!=\!0$ and $\dot{\mathcal{G}}\!=\!0$, and the time-independent condition can read from the Eq.~(\ref{R^n+G^m}), shown as
\begin{equation} \label{R^n+G^m time-independent}
    k_{1} R^n(\frac{n}{2}-1)= k_{2} \mathcal{G}^m (1-m).
\end{equation}
Because of the same order $H_0$, we have the relations
\begin{equation}  \label{R^n+G^m relation}
    n = 2m  \quad and \quad  -k_1 (-12)^{n} = k_2 (24)^{n/2}.
\end{equation}
We can take the Eq.~(\ref{R^n+G^m time-independent}) back to the Eq.~(\ref{R^n+G^m}), and we will find that the conserved quantity is always zero.

Checking this solution for $n\!=\!1$ and $m\!=\!1/2$, $f(R,\mathcal{G}) \!=\! k_1 R + k_1 \sqrt{6 \, \mathcal{G}}$, so the Eq.~(\ref{R^n+G^m}) changes as
\begin{equation}
    I = 6 a^3 H_{0}^2 k_{1} - 6 a^3 H_{0}^2 k_{1} = 0.
\end{equation}
Checking this solution for $n\!=\!2$ and $m\!=\!1$, $f(R,\mathcal{G}) \!=\! k_1 R^2 - 6k_1 \mathcal{G}$, so the Eq.~(\ref{R^n+G^m}) changes as
\begin{equation}
    I = - a^3 k_{1} R^2 + a^3 k_{1} R^2 = 0.
\end{equation}
We conclude that $I\!=\!I^{+} + I^{-}$ is always zero, due to the conserved quantity of positive $I^{+}$ and negative $I^{-}$ offset each other.
The conserved quantity term of $R$ provide for $I^{+}$, and the conserved quantity term of $\mathcal{G}$ provide for $I^{-}$. These show that the accelerated expansion cosmology of this case of $f(R,\mathcal{G})$ theory has only the conservation of energy, and the corresponding conserved quantity is zero, for the balance of the conserved quantity of positive and negative.
Finally, we can get the general form for available solution of the accelerated expansion cosmology, which is deduced as
\begin{equation}
   f(R,\mathcal{G}) = k_1 R^n + (6)^{\frac{n}{2}}(-1)^{n+1} k_1 \mathcal{G}^{\frac{n}{2}}. \label{case1}
\end{equation}
We can find that this solution can match the de Sitter solution of the formula (46) in the article~\cite{fRGBnoether}, and the exact coefficient is the same for $\frac{c_1 + 3c_2}{2c_1} = n$.

If we also consider the energy condition~\cite{fGBnoether}, corresponding to the 00 component of Einstein equation, which is shown as
\begin{equation} \label{energy condition00}
  \bigg(\frac{\dot{a}}{a} \bigg)^2 f_{R} + \bigg(\frac{\dot{a}}{a} \bigg) \frac{d\, f_{R}}{dt}  - 4 \bigg(\frac{\dot{a}}{a} \bigg)^3 \frac{d\,  f_{\mathcal{G}}}{dt} - \frac{1}{6}\bigg[f - R f_{R} - \mathcal{G} f_{\mathcal{G}} \bigg] = 0.
\end{equation}
For the de Sitter solution $a(t)\propto e^{H_{0}t}$, the Ricci scalar $R$ and Gauss-Bonnet invariant $\mathcal{G}$ change as $ R\!=\!-12 H_{0}^2$ and $ \mathcal{G}\!=\!24 H_{0}^4$, which mean $\dot{R}\!=\!0$ and $\dot{\mathcal{G}}\!=\!0$. Then the energy condition can be read from the Eq.~(\ref{energy condition00}), shown as
\begin{equation}
      k_{1} R^n(\frac{n}{2}-1)= k_{2} \mathcal{G}^m (1-m).
\end{equation}
Obviously, we find that the constraint solutions of $m$ and $k_2$ obtained by the energy condition~(\ref{energy condition00}) are the same as the Eq.~(\ref{R^n+G^m time-independent}) by the time-independent condition of the conserved quantity $\dot{I}\!=\!0$. It can be said that the conclusion, obtained from the symmetric conservation, is consistent with the energy condition. Indirectly, the correctness of this general symmetry research is proved.

\subsection{Case \uppercase\expandafter{\romannumeral2}: $f(R,\mathcal{G}) = k R^n \mathcal{G}^m $}
Under the premise of satisfying completeness of the commutator relations, the corresponding generalized Noether symmetric generators can be obtained as
\begin{eqnarray}
  X_{1} &=& \bigg(-3 F_1(t) + 1 \bigg)\frac{\partial}{\partial t} + \dot{F_1}(t) a\frac{\partial}{\partial a}, \\
  X_{2} &=& -3 F_1(t)\frac{\partial}{\partial t} + \dot{F_1}(t) a\frac{\partial}{\partial a}.
\end{eqnarray}
The generators are the same as the Case \uppercase\expandafter{\romannumeral1} $f(R,\mathcal{G}) = k_1 R^n + k_2 \mathcal{G}^m$, so the $\dot{F}_1(t)$ can only be zero, and the only remaining generalized noether symmetric generator and the corresponding first integral
\begin{eqnarray}
  X &=& \frac{\partial}{\partial t},  \nonumber \\
  I &=& 6 a^3 H k \dot{R} \frac{R^{n} \mathcal{G}^m}{R^2} (n^2-n) - 24 a^3 H^3 k \dot{\mathcal{G}} \frac{R^{n} \mathcal{G}^{m}}{\mathcal{G}^{2}} (m^2-m)  + a^3 k R^{n} \mathcal{G}^{m} (n+m-1)  \nonumber \\
    &+& 6 a^3 H^2 k \frac{R^{n} \mathcal{G}^m}{R} n - 24 a^3 H^3 k \dot{R} \frac{R^{n} \mathcal{G}^{m}}{R \mathcal{G}} nm + 6 a^3 H k \dot{\mathcal{G}} \frac{R^{n} \mathcal{G}^{m}}{R \mathcal{G}} nm . \label{I=R^nG^m}
\end{eqnarray}
If we choose the condition of $n+m\!=\!1$, then $f(R,\mathcal{G}) \!=\! k R^n \mathcal{G}^m$ will be reduced to $k R^n \mathcal{G}^{(1-n)}$. The corresponding symmetric generators change as
\begin{eqnarray}
  X_{1} &=& \bigg(1-3 F_1(t) \bigg)\frac{\partial}{\partial t} + \dot{F_1}(t) a \frac{\partial}{\partial a} + \frac{R}{\mathcal{G}}F_2(t,R,a+\mathcal{G})\frac{\partial}{\partial R} + F_2(t,R,a+\mathcal{G})\frac{\partial}{\partial \mathcal{G}}, \\
  X_{2} &=& -3 F_1(t)\frac{\partial}{\partial t} + \dot{F_1}(t) a \frac{\partial}{\partial a} + \frac{R}{\mathcal{G}}F_2(t,R,a+\mathcal{G})\frac{\partial}{\partial R} + F_2(t,R,a+\mathcal{G})\frac{\partial}{\partial \mathcal{G}}.
\end{eqnarray}
For the same condition $[X_1, X_2]\!=\!X_2$ or $[X_1, X_2]\!=\!X_1$, we have
\begin{equation}
    \ddot{F_1} = \dot{F_1} = F_1 = 0 , \quad \dot{F_2} = F_2 \quad and \quad  \dot{R} \mathcal{G} = R \dot{\mathcal{G}}.
\end{equation}
The corresponding symmetric generators can be simplified as
\begin{eqnarray}
  X_{1} &=& \frac{\partial}{\partial t} + \frac{R}{\mathcal{G}}F_2\frac{\partial}{\partial R} + F_2\frac{\partial}{\partial \mathcal{G}}, \\
  X_{2} &=& \frac{R}{\mathcal{G}}F_2\frac{\partial}{\partial R} + F_2\frac{\partial}{\partial \mathcal{G}},
\end{eqnarray}
which indicate the only conserved quantity, because of $I_2\!=\!0$, simplified as
\begin{equation}
   I_1 = 6 n a^3 H^2 k \bigg(\frac{\mathcal{G}}{R}\bigg)^{1-n}.
\end{equation}
Finally, we consider the constraints of the conservation condition $\dot{I_1}\!=\!0$ and $\dot{R} \mathcal{G} \!=\! R \dot{\mathcal{G}}$. We can classify the cosmology solutions according to the $m$,
\begin{align}
  n>1  && a(t)\propto t^{1/2}  &&  R=0              && f(R,\mathcal{G})\rightarrow f(\mathcal{G}), \\
  n<1  && a(t)\propto t        &&  \mathcal{G}=0    && f(R,\mathcal{G})\rightarrow f(R),
\end{align}
and it's not hard to find that we can contain the previous analysis into this general case. Both of the cosmology solution are not suitable to explain the accelerated expansion cosmology.

Back to the general conserved quantity Eq.~(\ref{I=R^nG^m}) of the case $f(R,\mathcal{G}) \!=\! k R^n \mathcal{G}^m $, if we don't choose the condition of $n+m\!=\!1$, we can use the similar approach to analyze the conditions for the de Sitter solution $a(t)\! \propto \! e^{H_{0}t}$. Furthermore, the corresponding conserved quantity changes as
\begin{equation}
     I = a^3 k R^n \mathcal{G}^m (\frac{n}{2}+m-1),
\end{equation}
which means that the time-independent condition should be
\begin{equation} \label{R^nG^m time-independent}
    \frac{n}{2}+m-1 = 0,
\end{equation}
then the general form $f(R,\mathcal{G}) \!=\! k R^n \mathcal{G}^m $ reduces to
\begin{equation}
    f(R,\mathcal{G}) = k \bigg(\frac{R}{\sqrt{\mathcal{G}}} \bigg)^n  \mathcal{G}. \label{case2}
\end{equation}
We can also find that this solution can match the de Sitter solution of the function (42) in the article~\cite{fRGBnoether}, and the exact coefficient is the same. In addition, we can determine the specific form of the function (42), shown as
\begin{eqnarray}
    f(R,\mathcal{G}) &=& R^2 \tilde{f}(\frac{\mathcal{G}}{R^2}) = \mathcal{G} \frac{R^2}{\mathcal{G}} \tilde{f}(\frac{\mathcal{G}}{R^2}) = \mathcal{G} \bar{f}(\frac{R^2}{\mathcal{G}}). \nonumber \\
    \bar{f}(\frac{R^2}{\mathcal{G}}) &=& k \bigg(\frac{R}{\sqrt{\mathcal{G}}} \bigg)^n .
\end{eqnarray}

If we consider the energy condition again, taking the de Sitter solution $a(t)\propto e^{H_{0}t}$ of the accelerated expansion cosmology into the Eq.~(\ref{energy condition00}), we find the solution is
\begin{equation}
      \frac{n}{2}+m-1 = 0.
\end{equation}
Obviously, the constraint solutions of $m$, obtained by the energy condition~(\ref{energy condition00}), is the same as the Eq.~(\ref{R^nG^m time-independent}) of the conservation condition. And again, we have proved that the generalized Noether symmetry includes the energy condition.

For the special case of $n\!=\!2$ and $m\!=\!0$, so $f(R,\mathcal{G})\! \Rightarrow \! R^2 $, or the case of $n\!=\!1$ and $m\!=\!1/2$, so $f(R,\mathcal{G}) \!\Rightarrow \! k R\sqrt{\mathcal{G}}$. These simple modification models have been discussed before. However, there are many models of the different $n$ need to detailed study, all of these contain the accelerated expansion solution. And we can determine the accurate value of $n$ by dynamic stability analysis and the constraints of cosmological observation data. Noting that the corresponding conserved quantity is still zero, and the only conservation relation is the conservation of energy.

\section{Conclusion}
\label{conclusion}

In the extended gravity theories, the explicit expressions of Lagrangian will determine the exact forms of solutions of the field equation. It is important to analyze physical properties of the theories. Meanwhile, the generalized Noether's theorem is a useful method to define the form of a theory and obtain physical explanations such as the conserved quantities and symmetries. In this paper, a theory of $f(R,{\mathcal{G}})$ gravity is discussed, whose interaction Lagrangian consists of the Ricci scalar $R$ and the Gauss-Bonnet invariant $\mathcal{G}$. After some considerations based on cosmological evolution from this model, the generalized Noether symmetries are investigated. The methodology used in this work can also be adopted for other modified gravity theories. It is unavoidable to modify the expressions similar to what we have done in this paper.

Although we can not deduce the concrete forms directly, we can get the polynomial and product forms of $f(R,\mathcal{G})$ theories satisfying the generalized Noether symmetries, and obtain the meaningful explanations of the corresponding conserved quantities and symmetries. Both forms of $f(R,\mathcal{G})$ concerned in this paper only possess the time translational symmetry. We obtain the general solutions of polynomial and product forms. Then we find that the constraint condition $\dot{R} \mathcal{G} \!=\! R \dot{\mathcal{G}}$, simplifying the forms of conserved quantities, is a strong and convenient condition. But this convenience also inevitably constrains the forms of solutions. This difficulty may be solved by calculating the Hojman symmetry~\cite{Hojman_symmetry} using the equations of motion, rather than the generalized Noether symmetry using the (point-like) Lagrangian. This research topic will be done in our future work.

Fortunately, through some calculations for the solution of the late-time accelerated expansion cosmology, we obtain two general forms.
Both of them can be satisfied with the energy condition. To some extent, we have demonstrated the consistency of the conservation of symmetry and the energy condition. One is the polynomial form $f(R,\mathcal{G}) \!=\! k_1 R^n + (6)^{\frac{n}{2}}(-1)^{n+1} k_1 \mathcal{G}^{\frac{n}{2}}$, and the other is the product form $f(R,\mathcal{G}) \!=\! k ( R / \sqrt{\mathcal{G}} )^n \mathcal{G}$.
Both of them are part of the article~\cite{fRGBnoether}, because we consider directly the de Sitter solution for the accelerated expanding of our universe, and there's no contradiction here. Although this is only a partial result, our results are more specific and the coefficients are clearer.
The corresponding conserved quantities are both zero, and the only conservation relation is the conservation of energy. It seems that we are hard to observe other conserved quantities and symmetries. For the concrete form of different $n$, we will get some interesting solutions by the dynamic stability analysis in our further paper. And it can also be restrained by the cosmological observation data. Finally, it is worth emphasizing that symmetry is not only a mathematical tool for solving dynamic problems, but also allows for the selection of physically observable universes, especially analytical models related to observations as discussed in~\cite{dis Noether}.

\section*{Acknowledgements}

This work has been supported by the Young and Middle-aged Teachers Education Research Project of Fujian Province under the Grant No. JT180738 and the Fuzhou Institute of Technology.

\end{document}